\documentclass[aps,prl,groupedaddress,twocolumn,showpacs,superscriptaddress]{revtex4-1}

\usepackage{graphicx}
\usepackage{amsmath,amssymb,amsfonts}

\begin{document}


\title{Wave-packet numerical investigation of thermal diffuse scattering:\\
A time-dependent quantum approach to the Debye method}

\author{S. Rudinsky}
\affiliation{Department of Mining and Materials Engineering, McGill
University, 3610 University Street, Montreal, Qc, Canada, H3A 0C5}

\author{A. S. Sanz}
\affiliation{Department of Optics, Faculty of Physical Sciences,
Universidad Complutense de Madrid,\\
Pza.\ Ciencias 1, Ciudad Universitaria 28040 - Madrid, Spain}

\author{R. Gauvin}
\affiliation{Department of Mining and Materials Engineering, McGill
University, 3610 University Street, Montreal, Qc, Canada, H3A 0C5}

\begin{abstract}
The effects of thermal diffuse scattering on the transmission and eventual diffraction of highly accelerated electrons are investigated with a method that incorporates the frozen phonon
approximation to the exact numerical solution of the time-dependent Schr\"odinger equation.
Unlike other methods in the related literature, in this approach the attenuation of diffraction
features arises in a natural way by averaging over a number of wave-packet realizations, thus
avoiding any additional experimentally obtained Debye-Waller factors or artificial modulations. Without loss of generality, the method has been applied to analyze the transmission of an electron
beam through a thin Al film in two dimensions, making use of Einstein's model to determine the phonon
configuration for each realization at a given temperature.
It is shown that, as temperature and hence atomic vibration amplitudes increase, incoherence among
different electron wave-function realizations gradually increases, blurring the well-defined diffraction
features characterizing the zero-temperature intensity.
\end{abstract}

\date{\today}


\maketitle


Experiments are often performed at room temperature causing temperature driven atomic vibrations or thermal diffuse scattering (TDS) to
importantly affect the acquired data.
Consequently, any methodology aimed at analyzing such data has to be devised in such
a way that it can properly reproduce the effects of thermal vibrations.
Currently, two major techniques are used in this regard,
namely the Debye method and the frozen phonon model, both of which are applied to studies ranging from determining mechanical properties to electron imaging
\cite{SASAKI2016240,Winkelmann2007414,wang1995elastic}.
In the case of the frozen phonon model, current approaches are based on performing
wave function propagations by invoking the paraxial approximation, which constrains
the simulations to applications in the high-energy regime \cite{kirkland2010advanced}.
However, for applications within the low-energy regime,
such an assumption is no longer valid and simulations require
the use of time-dependent approaches.
As of now, computations of phonon excitation including TDS for time-dependent
wave packet propagation have not been thoroughly investigated.

In this Letter we tackle the issue by exactly solving the electron time-dependent
Schr\"odinger equation, avoiding the many-body problem associated with atomistic
motions by appealing to the frozen phonon model.
It is shown that attenuation of diffraction peaks naturally arises without the
use of smoothing functions and experimentally obtained Debye-Waller factors.
The Debye method adjusts the averaged
material potential of a crystal using a combination of the
Debye-Waller factor \cite{peng1999electron,GilatPhysRev143} and
another portion to account for the TDS
\cite{NicklovPhysRev1966Lattice},
\begin{equation}
\label{eq:DWandTDS}
 I_{\mbox{obs}}=Ie^{-2M}+TDS ,
\end{equation}
where $M$ is the Debye-Waller factor and $I$ is the intensity of a
stationary, non-vibrating lattice. The Debye-Waller factor itself is
empirically determined from experimental data obtained by X-ray or neutron
diffraction \cite{McDonald:a05638,PRYOR1964275,Chipman1960DW}. The
TDS portion of Eq.~\ref{eq:DWandTDS} is often modeled by the use of
complex potentials which act as absorbers \cite{Peng:zh0008}. While
the Debye method can be used to obtain diffraction peaks affected by
TDS, multiple scattering cannot be reproduced using absorbing
potentials because once a portion of the wave function is absorbed,
it is removed from the total probability density. This also makes it
impossible to ensure measured quantities, such as the energy of the
system are conserved \cite{AMULLER2001371}.

\begin{table*}[!t]
\caption{Vibration amplitudes and corresponding temperatures, according to Eq.~(\ref{u2-T-rel}),
 considered here for Al(100).}
 \begin{tabular}{c|c|c|c|c|c|c|c|c|c|c}
 \hline\hline
  $\sqrt{\langle \| {\bf u}\|^2\rangle}$ (\AA) & 0.0265 & 0.0529 & 0.0794 & 0.106 & 0.132 & 0.159 &0.185& 0.212 & 0.238 & 0.265\\ \hline
  $T$ (K) & 6.7 & 26.7 & 60.2 & 107.3 & 166.4 & 241.4 &327.4& 429.19 &541.2& 670.6\\
  \hline\hline
 \end{tabular}
 \label{tab:rvsT}
\end{table*}

To properly describe the atomic vibrations and hence TDS,
Einstein's model is incorporated in the frozen phonon approach, often used
in calculations such as diffraction image simulations \cite{Allen201511}.
The method assumes that each electron ``feels'' a single frozen atomic configuration.
Treating phonon excitations as both independent \cite{Loane:st0492} and correlated
\cite{Jesson273,AMULLER2001371} harmonic oscillators has been
thoroughly explored, showing that Einstein's model is sufficient to obtain general
information about diffraction patterns, while correlated phonon excitations are
necessary to distinguish small characteristics of complex patterns in numerical
simulations \cite{AMULLER2001371}.
Typically, a phonon
configuration is first chosen at random, with the atomic displacements from equilibrium
following a Gaussian distribution.
Then, the electron wave function is propagated in time to obtain the exit wave function
and diffraction pattern \cite{kirkland2010advanced,LOANE1992121}, which constitutes a
single, non-dissipative realization.
After performing a large number of such simulations with different random phonon
configurations at the same energy, Monte Carlo integration of the
detected wave function provides a simulated diffraction pattern
including TDS effects
\cite{Allen201511,kirkland2010advanced,Loane:st0492,AMULLER2001371,FINDLAY2005126}.
The nature of this method makes simulating plural scattering
possible and permits quantification of observables.

More specifically, thermal vibrations have been incorporated by
displacing each atom about its mean position following a Gaussian
probability distribution. The probability of the $i$th atom being in
position ${\bf r}_i$ is given by
\begin{equation}
 P({\bf r}_i) = e^{-\|{\bf r}_i - {\bf r}_0\|^2/2 \langle \|{\bf u}\|^2\rangle} ,
 \label{eq:pdf}
\end{equation}
where ${\bf r}_0$ is the mean position of the $i$th atom and
$\langle \| {\bf u}\|^2\rangle$ the root mean squared vibration amplitude.
The vibration amplitude is related to the Debye-Waller factor by
\begin{equation}
\label{eq:Bfromu}
 M = 8 \pi^2 \langle \| {\bf u}\|^2\rangle ,
\end{equation}
and can be related to the temperature of the material by the
following expression \cite{Peng:zh0008,Cartz1955Thermal},
\begin{equation}
 \sqrt{\langle u^2\rangle}=7.816\times10^{-2}a\sqrt{\frac{T}{T_M}} ,
 \label{u2-T-rel}
\end{equation}
where $a$ is the stationary lattice parameter and $T_M$ is the
melting temperature. Once the atom positions are determined, the
average potential of the material is calculated using the
parametrization of scattering factors given by Peng \cite{peng1999electron}.
In our case, the potential simulates a thin Al(100) film in two dimensions (2D).

Once the potential model is set, the numerical simulation of the electron
transmission has been carried out according to the method described in
earlier work for fully coherent scattering \cite{RudinskyNearfield2016}.
The vibration amplitudes investigated and the corresponding material
temperatures are displayed in Table~\ref{tab:rvsT}.
The initial state of each electron has been modeled as a Gaussian wave
packet with spread $\sigma_x = a$ and $\sigma_y=0.5a$. The initial
wave function is positioned above the thin film and propagation is
performed in the $z$-direction normal to the material surface. This
configuration has been chosen to closely represent an electron probe. The
electron probe of an electron microscope is generated by eliminating
frequencies from a plane wave which are outside a specified range.
As a result, the incident wave function is a Gaussian wave packet at
the sample plane. The kinetic energy of the initial wave function in
all cases is 1~keV, so the wave packet moves towards the film with a
speed of $1.88\times10^6$~m/s.
Since the thickness of the film chosen to be in our calculations is 80~\AA\, it takes the electrons about 4.26~fs to pass through this material.
This time is much faster than the typical vibration period, which is of the
order of picosecond (the phonon frequencies are within the range
$10^{12}$--$10^{13}$~Hz \cite{kirkland2010advanced}, thus validating
the frozen phonon model here considered.

\begin{figure}[!b]
 \includegraphics[width=8cm]{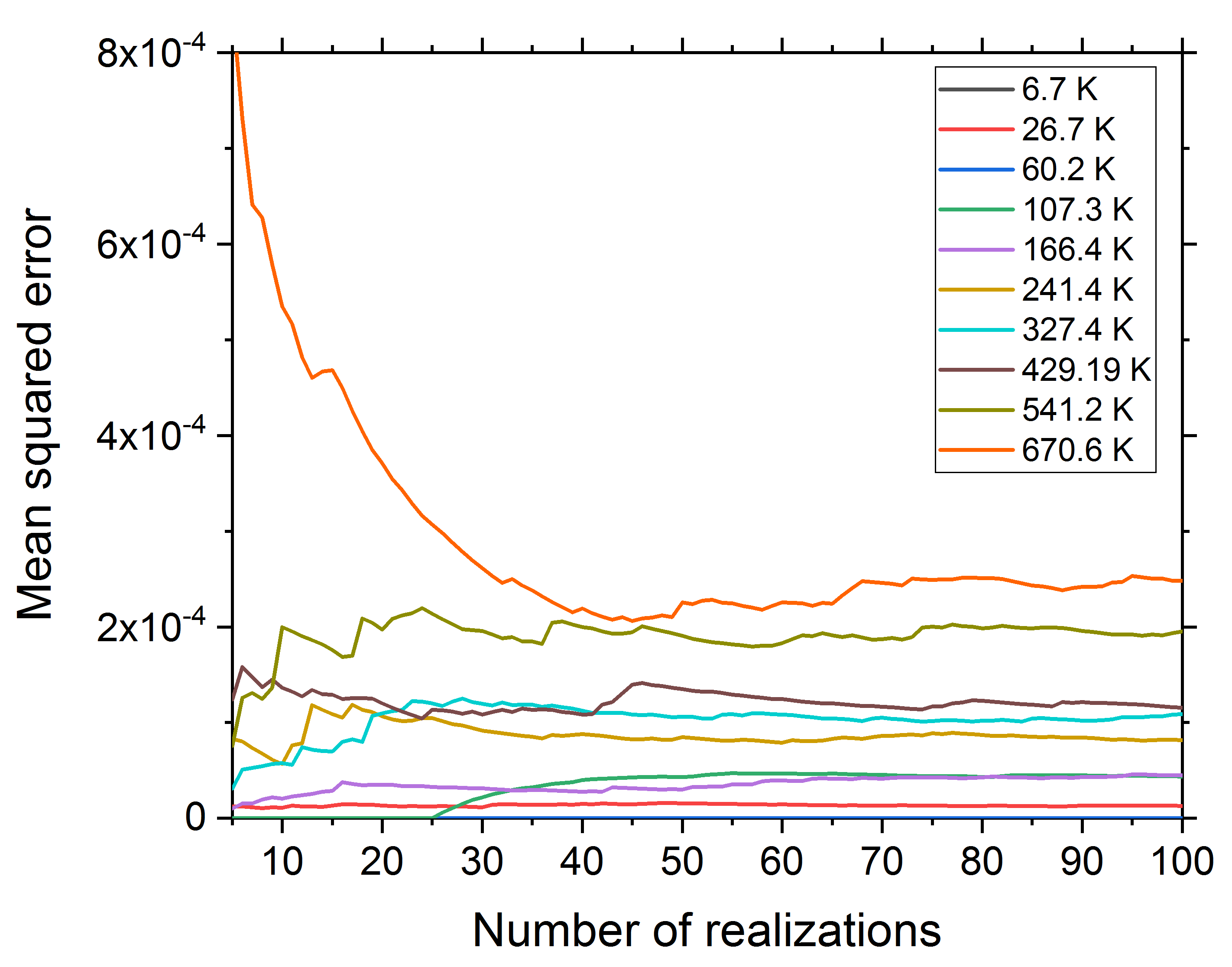}
 \caption{Average MSE for each of the temperatures in Table~\ref{tab:rvsT}.
  Beyond approximately $N=50$ realizations all simulations are
  well converged (the average MSE is nearly constant).}
 \label{fig:MSE}
\end{figure}

Wave packet propagation has been performed by solving the time-dependent
Schr\"{o}dinger equation numerically using the split-operator method
\cite{FEIT1982412,LEFORESTIER199159}. For each temperature, 100
realizations have been performed, each with a different random
atomic configuration obtained by following the above mentioned
method.
Each realization is finalized once the wave packet has crossed the material
and the interaction potential affecting it is negligible.
To ensure that there are no unphysical reflections at the boundaries of the
numerical grid, an absorbing function has been considered, which removes
possible outgoing flow without affecting the correct evolution of the wave-packet
in the region of interest \cite{Kosloff1986363}. Intensity distributions as a function
of the parallel momentum transfer and diffraction patterns in momentum space
have then been calculated by averaging the values at each
grid point over all the realizations performed. In order to ensure
convergence of the method, the mean squared error per grid
point of the diffraction pattern,
\begin{equation}
\label{eq:MSE}
 \mbox{MSE}(p) = \frac{1}{N}\sum_{i=1}^N [X_i(p) -\bar{X}(p)]^2 ,
\end{equation}
has been calculated at each temperature for a number of realizations ($N$) ranging
between 5 and 100.
In Eq.~(\ref{eq:MSE}), $X_i$ is the value at the grid point $p$ in the diffraction
pattern of the $i$th realization and $\bar{X}$ is the value at the corresponding point
in the average diffraction pattern. An average
of the MSE has then been taken over all the grid points to
obtain a single value. The results for each vibration amplitude shown in Table~\ref{tab:rvsT} are
shown in Fig.~\ref{fig:MSE}.


\begin{figure}
 \includegraphics[width=8cm]{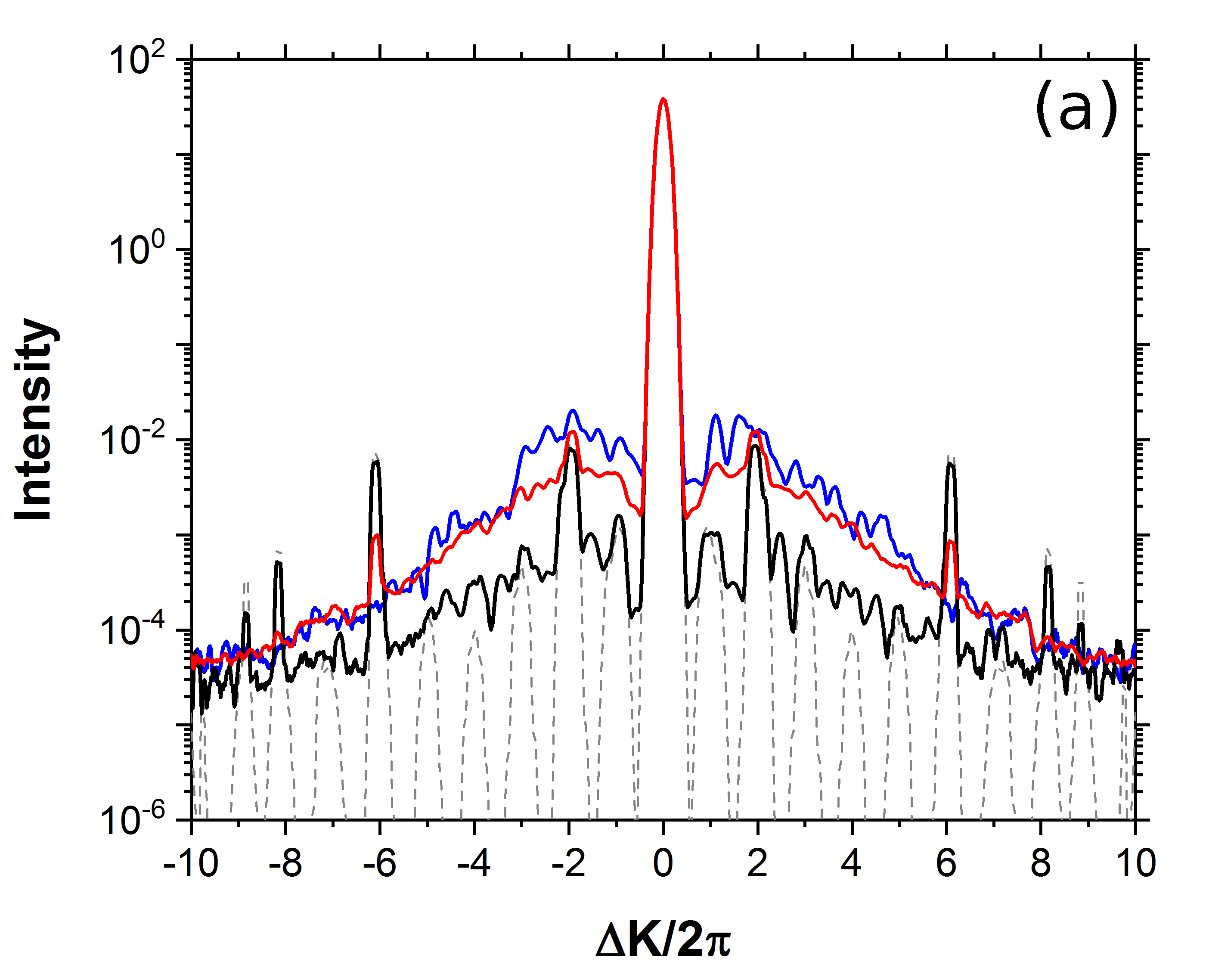}
 \includegraphics[width=8cm]{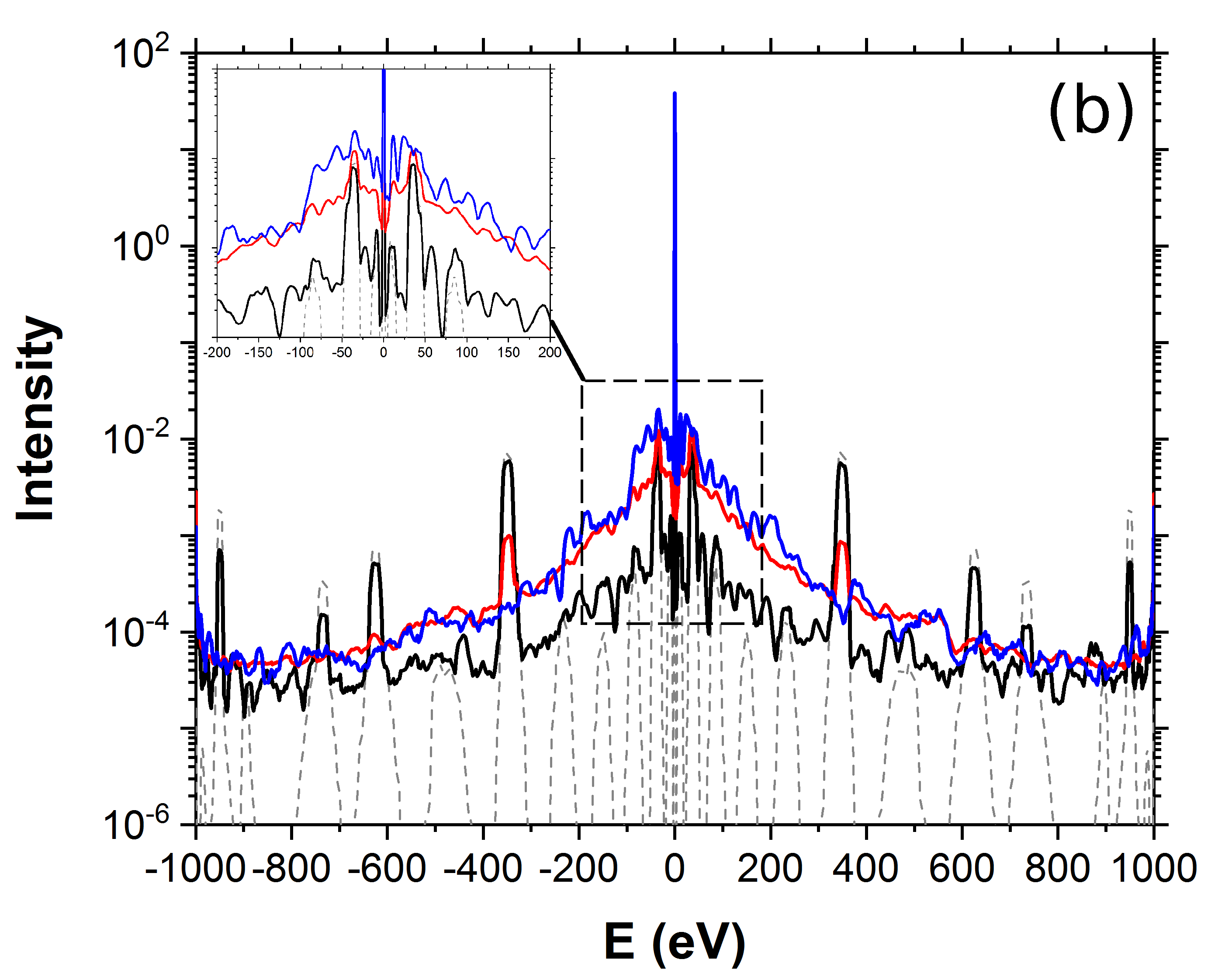}
 \caption{Intensity distribution as a function of (a) the parallel momentum and (b) the energy transferred to the
  film at different temperatures: 0~K (dashed line), 26.7~K (black solid line), 241.4~K (red solid line), and
  670.6~K (blue solid line).}
 \label{fig:smatrix}
\end{figure}

The intensity distribution as a function of the parallel momentum
transferred to the film is shown in Fig.~\ref{fig:smatrix}. The peak
intensities for each wave vector can be clearly seen in the case of
a static crystal.
However, as temperature increases, the intensity distribution becomes
increasingly diffuse.
Notice that, because the simulations are performed at a relatively low energy,
the scattering probability is high and incoherence can be easily seen even for
such a thin specimen.
Nevertheless, the crystal structure is not completely lost even at high temperatures,
where we can see the persistence of the diffraction peaks for the orders $\pm 2$,
$\pm 6$, $\pm 8$, and $\pm 9$, which are the most prominent ones in the zero temperature case.
As would be expected, as temperature increases, they become relatively weaker and weaker.
At $T = 241.4$~K, for instance, only the $\pm 2$ and $\pm 6$ orders remain, and at
$T = 670.6$~K only kind of broad shoulders around $\pm 2$ remain.

\begin{figure}
\includegraphics[width=7cm]{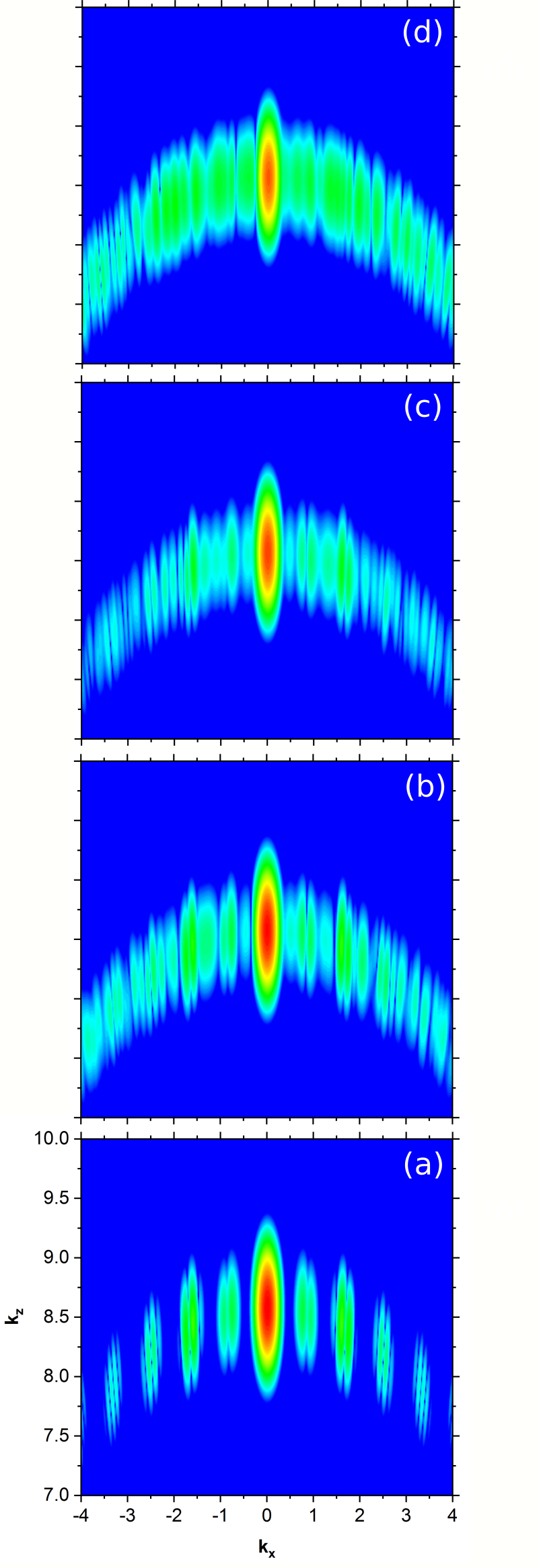}
 \caption{Averaged diffraction densities in the momentum representation obtained
  from averaging over 100 simulations at different temperatures: (a) 0~K, (b) 26.7~K,
  (c) 241.4~K, and (d) 670.6~K.}
 \label{fig:DiffPat}
\end{figure}

The intensity displayed in Fig.~\ref{fig:smatrix} is a reflection of the dynamics
described by the associated density in the momentum representation.
Asymptotically, the influence of the potential is negligible on the wave function,
even if spatially it is still well inside the near field and does not display any
of the recognizable features of diffraction patterns.
However, from the momentum perspective it is already well converged and we can
extract valuable information from its structure without going to the far
field.
Bearing this in mind, we have computed the diffraction patterns in the momentum
representation for each vibration amplitude and compared them to that of the
static crystal.
Plots are displayed in Fig.~\ref{fig:DiffPat}, in particular, from bottom to
top, for 0~K, 26.7~K, 241.4~K, and 670.6~K.
As can be seen, the effects of TDS are already quite clear.
When TDS is not incorporated into the model, clear diffraction spots
unique to the crystal structure can be distinguished. However, as
temperature and, consequently, vibration amplitude increase, such
spots become less distinguishable and the resulting pattern is
diffuse, resembling that of an amorphous material. In these types of
materials, disorder in the atomic planes causes incoherent
scattering of the wave function inside the material
\cite{williams2009transmission}. In the case of the frozen phonon
model, integrating over multiple different configurations replicates
this incoherence \cite{LOANE1992121}.

Figures~\ref{fig:smatrix} and
\ref{fig:DiffPat} show that the frozen phonon model applied to
time-dependent and low energy wave packet propagation reproduces TDS
without the need of artificial modulation by Debye-Waller factors
and complex potentials. This is because decreases in peak
intensities farther from the primary beam and diffusivity of
intensity between peaks arise naturally through this method.
Integrating over many phonon configurations results in an exit wave
function which seems to have interacted with a smeared potential,
whose degree of smearing is a function of temperature. This
simulates incoherent scattering of the wave function, which removes
intensity from the portions which would otherwise be coherently
scattered. Therefore, these peaks are less intense and intensities
from angles between those of the wave vectors are present in the
form of noise.

\begin{figure}[!t]
 \includegraphics[width=8cm]{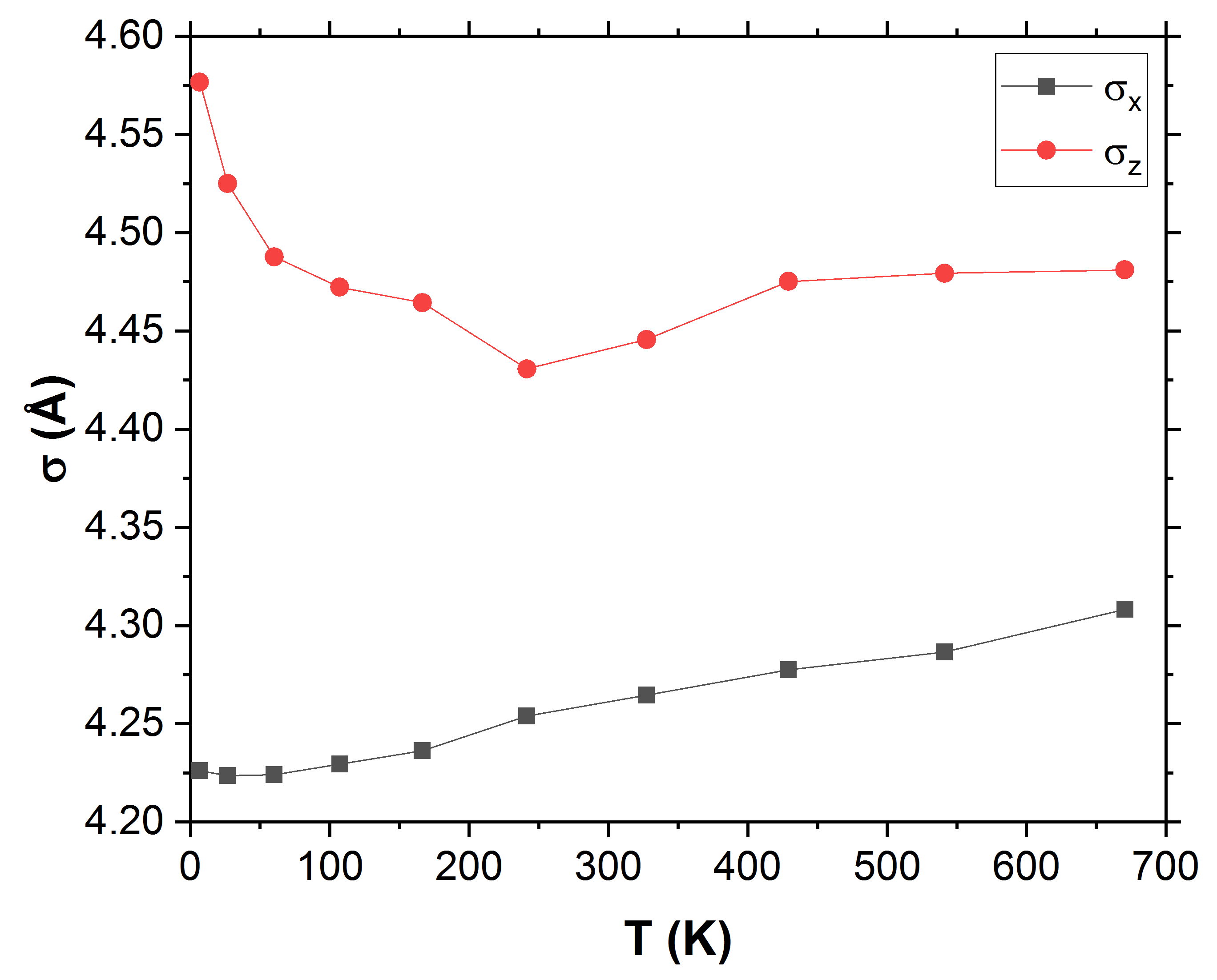}
 \caption{Spatial dispersion as a function of temperature for wave-packet propagation including TDS.}
 \label{fig:dxvT}
\end{figure}

In our approach, for each single realization, the transmission process only involves elastic
scattering regardless of the temperature.
To confirm this fact, both the expectation value of the energy, $E = \langle \hat{H} \rangle$, and its
dispersion,
\begin{align}
\label{eq:dE}
\Delta E &= \sqrt{\langle \hat{H}^2\rangle-\langle \hat{H}\rangle^2} ,
\end{align}
have been calculated.
As expected, the expectation value of the energy for each
temperature has remained equal to that of the incident particle, 1~keV; the same has been
observed for the energy dispersion, which has remained equal to 0.054~keV.
On the other hand, it is well known \cite{Gauvin201621,Clifflorimer1981scattering}
that when the wave packet is traveling through a material, it can also be affected
by the number of scattering events occurring between the incident particle and the surrounding atoms,
which will also be dependent on the thermal agitation of the latter.
To quantify this effect, we have also computed the spatial dispersion along the $x$ and
$z$ directions,
\begin{equation}
\label{eq:dx}
 \sigma_s = \sqrt{\langle\hat{s}^2\rangle-\langle\hat{s}\rangle^2}
\end{equation}
where $s = x, z$, averaging it in the same way as the energy.
The results as a function of temperature for the
simulations done in this work are presented in Fig.~\ref{fig:dxvT}.
It is interesting to note that while the dispersion along the propagation direction is an
increasing function (nearly linear) with temperature, along the transverse direction
it decreases until approximately 250~K and then undergoes a very slight increase, which becomes
constant for large temperatures.

In the case of the static lattice, it was shown previously that for
time-dependent wave packet propagation through an 80~\AA\ thick Al
film with the same initial conditions, broadening in the lateral
direction was unaffected by the material and followed that of a free
particle \cite{RudinskyNearfield2016}. Figure~\ref{fig:dxvT} shows
that in fact with the incorporation of TDS, lateral spreading
increases with increasing temperature while longitudinal spreading
decreases for an initial incorporation of disorder and then remains
relatively constant. The variation in the $x$-direction implies that
scattering from an imperfect crystal increases beam broadening. In
the above mentioned previous work, it was shown that upon exit of
the material, the spread of the wave function was 1.02 times its
initial size \cite{RudinskyNearfield2016}. Here, the spread of the
final wave function varies from 1.055$\pm0.004$~\AA\ to 1.08$\pm0.02$~\AA.
This increase in spread corresponds to the diffusivity and increased
amount of high intensity regions between the diffraction spots of
Fig.~\ref{fig:DiffPat}. In slicing methods, a 2D transmitted wave
function is computed at each slice so that full transmission in $z$
is assumed and the rays are approximately parallel to the incident
beam. Consequently, modulations in all spatial coordinates are
unobtainable. Here, a time-dependent approach demonstrates that
variations of the wave function in all dimensions occur with TDS and
that broadening is increased with disorder of the crystal.


While computing particle diffraction intensities, such as in X-ray
diffraction or electron microscopy, it is important to take into
account the effect of thermal diffuse scattering by phonon
excitations. In material properties calculations, such as
thermodynamic properties, the Debye method is typically used to
modulate the peak intensities while a constant complex potential
term is included in the real potential to increase background. In
transmission electron microscopy, the frozen phonon model is a
common tool where averaging over various different phonon
configurations reproducing the effects TDS will have on the
transmission image. This method is commonly used in high energy
modeling techniques but has not been applied to time-dependent,
low-energy situations. Here, the frozen phonon model has been
applied to a time-dependent solution of the Schr\"{o}dinger equation
of an electron beam transmitted through a thin aluminum film.
Incoherent scattering was accurately reproduced in the computed
intensity distributions and diffraction patterns by means of diffuse
intensities about the Bragg spots. It was shown that the method
conserves energy and that the spatial dispersion increases laterally
with increasing temperature due to an increased amount of scattering
events. Overall, such a method is not only reserved to slicing
techniques and the physical consequences of phonon excitations can
be reproduced without artificial modulation of diffraction
intensities.



S.R.\ and R.G.\ would like to acknowledge the Aluminum group (REGAL) for their financial support.
A.S.\ acknowledges financial support from the Spanish MINECO (grant No. FIS2016-76110-P).



\end{document}